\documentclass[journal]{IEEEtran}
\ifCLASSINFOpdf
  \usepackage[pdftex]{graphicx}
  \DeclareGraphicsExtensions{.pdf,.jpeg,.png}
\else
  \usepackage[dvips]{graphicx}
\fi
\usepackage{epstopdf}
\usepackage[cmex10]{amsmath}
\usepackage{amssymb}
\usepackage{wasysym}
\usepackage{algorithm}
\usepackage{algorithmic}
\usepackage{array}
\usepackage{cite}
\usepackage{color}
\usepackage{url}

\usepackage{epsfig,latexsym}
\usepackage{cite}
\usepackage{verbatim}
\usepackage{amsopn}
\usepackage{booktabs}
\usepackage{bm}
\usepackage{stfloats}
\usepackage{hyperref}
\usepackage{graphicx}
\usepackage{float}
\usepackage{subfigure}
\usepackage{makecell}
\usepackage{xcolor}
\usepackage{xpatch}

\begin{document}

\title{Joint Transmit Waveform and Receive Filter Design for ISAC System with Jamming}
\author{Yuan~Shu,~\IEEEmembership{Graduate~Student~Member,~IEEE}, Chenhao~Qi,~\IEEEmembership{Senior~Member,~IEEE}, Shiwen Mao,~\IEEEmembership{Fellow,~IEEE} \\
\thanks{Copyright (c) 2025 IEEE. Personal use of this material is permitted. However, permission to use this material for any other purposes must be obtained from the IEEE by sending a request to pubs-permissions@ieee.org.} 
\thanks{This work was supported in part by the National Natural Science Foundation of China under Grant U22B2007. (\textit{Corresponding author: Chenhao~Qi}.)}
\thanks{Yuan~Shu and Chenhao~Qi are with the School of Information Science and Engineering, Southeast University, Nanjing 210096, China (e-mail: \{shuyuan,qch\}@seu.edu.cn).} 
\thanks{Shiwen Mao is with the Department of Electrical and Computer Engineering, Auburn University, Auburn, AL 36849-5201, USA. (e-mail: smao@ieee.org).}
}

\markboth{Accepted By IEEE TRANSACTIONS ON VEHICULAR TECHNOLOGY}
{}

\maketitle

\begin{abstract}
  In this paper, to suppress jamming in the complex electromagnetic environment, we propose a joint transmit waveform and receive filter design framework for integrated sensing and communications (ISAC). By jointly optimizing the transmit waveform and receive filters, we aim at minimizing the multiuser interference (MUI), subject to the constraints of the target mainlobe, jamming mainlobe and peak sidelobe level of the receive filter output as well as the transmit power of the ISAC base station. We propose two schemes to solve the problem, including joint transmit waveform and matched filter design (JTMD) and joint transmit waveform and mismatched filter design (JTMMD) schemes. For both schemes, we adopt the alternating direction method of multipliers to iteratively optimize the transmit waveform and receive filters, where the number of targets as well as the range and angles of each target can also be estimated. Simulation results show that both the JTMD and JTMMD schemes achieve superior performance in terms of communication MUI and radar detection performance.
\end{abstract}

\begin{IEEEkeywords}
Alternating direction method of multipliers (ADMM), integrated sensing and communications (ISAC), mismatched filter, waveform design.
\end{IEEEkeywords}

\section{Introduction}
Integrated sensing and communications (ISAC), as a key technology for the sixth-generation (6G) wireless standard, has attracted wide interest from both the academia and industry~\cite{wei2023integrated,liu2022integrated,kaushik2024toward,meng2023throughput}. Different from traditional approaches that design communications and sensing separately, ISAC can share the hardware and wireless resources to perform radar sensing and wireless communications simultaneously and can achieve mutual benefits.

One crucial challenge for ISAC is waveform design. Radar prefers constant waveform for target detection, while the randomness of communication symbols causes the communication waveform to be time-varying. Therefore, the dual-functional waveform design needs to balance the different requirements of sensing and communications. Various design metrics have been adopted to meet different design goals~\cite{bazzi2023integrated,liu2021dual,wang2024robust}. For example, by minimizing the multiuser interference (MUI), the ISAC waveform is designed subject to the constraints of peak-to-average power ratio and the similarity constraint to a radar chirp signal~\cite{bazzi2023integrated}. The ISAC waveform is designed through minimizing the beampattern matching error subject to the communication quality-of-service requirements~\cite{liu2021dual}. More recently, the robust waveform design problem is considered, where the waveform is optimized by minimizing the MUI subject to the radar beampattern constraint and the total transmit power constraint~\cite{wang2024robust}. 

However, the aforementioned radar metrics, such as the similarity constraint to a radar chirp signal, do not necessarily ensure satisfactory performance for target detection. Therefore, we adopt the mainlobe level and peak sidelobe level as the radar performance metric. On the other hand, most of the existing works consider ISAC in an ideal electromagnetic environment and do not take radar jamming into consideration. In fact, the interrupted sampling repeater jamming (ISRJ) is a crucial and widely used radar jamming in modern radar warfare~\cite{wang2007mathematic}. The ISRJ can generate a series of false targets, seriously degrading radar performance and thus affecting the ISAC overall performance. To suppress the ISRJ, transmit waveform design and receive filter design are the two main aspects. By optimizing the orthogonal phase coding waveform, an improved genetic algorithm is proposed to mitigate the ISRJ~\cite{cao2021optimal}. Through exploiting the different time-frequency features of the ISRJ and target echo signal, the receive filter is designed to separate the ISRJ signal from the original signal~\cite{chen2019band}. To further enhance the anti-jamming performance, joint transmit waveform and receive filter design has been considered for radar systems, using alternating direction method of multipliers (ADMM)~\cite{zhou2020joint}, majorization minimization~\cite{gao2023joint}, or gradient-based nonlinear programming
together with the Lagrange multiplier method~\cite{wang2022complementary}. To the best knowledge of the authors, joint transmit waveform and receive filter design with the ISRJ has not yet been considered for the ISAC system.

In this paper, to suppress jamming in the complex electromagnetic environment, we propose a joint transmit waveform and receive filter design framework for the ISAC system. By jointly optimizing the transmit waveform and receive filters, we aim at minimizing the MUI, subject to the constraints of the target mainlobe, jamming mainlobe and peak sidelobe level of the receive filter output, as well as the transmit power of the ISAC base station (BS). We propose two schemes to solve the problem, including joint transmit waveform and matched filter design (JTMD) and joint transmit waveform and mismatched filter design (JTMMD) schemes. For both schemes, we adopt the ADMM to iteratively optimize the transmit waveform and receive filters, where the number of targets as well as the range and angles of each target can also be estimated. 

\textit{Notations: }Symbols for matrices and vectors are denoted in boldface, i.e., $m, \bm{m}$ and $\bm{M}$ denote a scalar, a vector and a matrix, respectively. $(\cdot)^{\rm T}, (\cdot)^{\rm H}, \|\cdot\|_2$ and $\|\cdot\|_{\rm F}$ represent the transpose, the conjugate transpose, the $\ell_2$-norm, and the Frobenius norm, respectively. $\circledast$ and $\odot $ denote the convolution operation and the Hadamard product, respectively. $\bm M[:,n]$ represents the $n$th column of the matrix $\bm M$. $\bm m[n]$ represents the $n$th entry of the vector $\bm m$. $\mathcal{CN}(\bm{m}, \bm{R})$ denotes the complex Gaussian distribution whose mean is $\bm{m}$ and covariance matrix is $\bm{R}$. $\mathbb{C}$ represents the set of complex-valued numbers. 

\section{System Model}\label{sec.system.model}
\begin{figure}[!t]
	\centerline{\includegraphics[height=6cm]{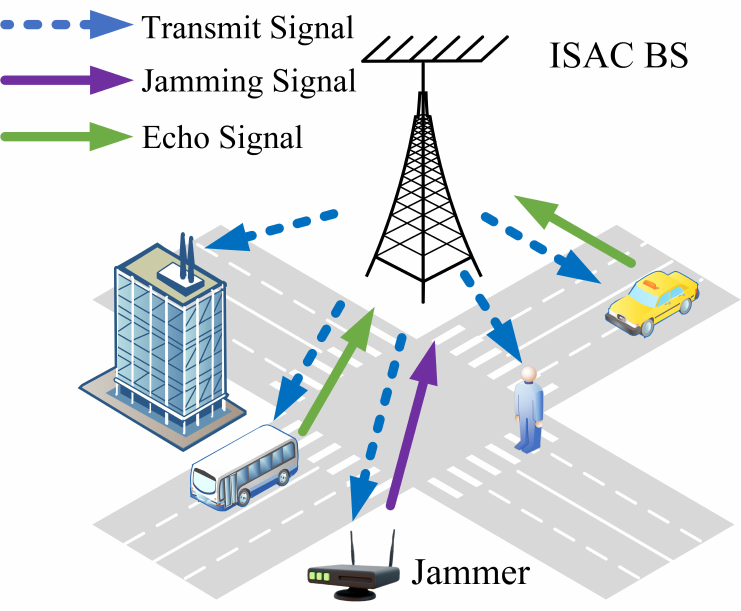}}
	\caption{Illustration of ISAC system with jamming.}
	\label{SystemModel}
\end{figure}

As shown in Fig.~\ref{SystemModel}, we consider an ISAC system, where an ISAC BS serves $M$ single-antenna communication users and senses $W$ targets simultaneously. The ISAC BS is equipped with $N_{\rm t}$ transmit antennas and $N_{\rm r}$ receive antennas. The antennas are placed in uniform linear arrays with half-wavelength intervals. Note that although this work focuses on narrow-band waveform design for the ISAC system, the proposed schemes can be readily extended to wideband scenarios, e.g., the OFDM framework.

For wireless communications, the received signal by the $M$ users can be expressed as
\begin{equation}
	\bm{Y}_{\rm c}=\bm{H}\bm{X}+\bm{N}_{\rm c},
\end{equation}
where $\bm{Y}_{\rm c}~\in\mathbb{C}^{M \times P}$ is the received signal, $\bm{X}~\in\mathbb{C}^{N_{\rm t} \times P}$ is the ISAC transmit waveform, and $P$ denotes the number of time slots. The communication channel $\bm{H}~\in\mathbb{C}^{M \times N_{\rm t}}$ is flat Rayleigh fading which remains constant within one communication frame and we assume that $\bm{H}$ is accurately estimated by the ISAC BS based on pilot symbols. $\bm{N}_{\rm c}\triangleq[\bm{n}_{{\rm c},1},\ldots,\bm{n}_{{\rm c},P}]\in\mathbb{C}^{M \times P}$ is the additive white Gaussian noise (AWGN), with $\bm{n}_{{\rm c},p}\sim\mathcal{CN}(\bm{0},\sigma_{\rm c}^2\bm{I}_{M}),~\forall p\in \mathcal{P}\triangleq\{1,2,\ldots,P\}$. 

Given the desired constellation symbol matrix $\bm{S}\in\mathbb{C}^{M \times P}$, we can rewrite the received signal as
\begin{equation}\label{MUI}
	\bm{Y}_{\rm c}=\bm{S}+\bm{\varXi}+\bm{N}_{\rm c},  
\end{equation}
where 
$\bm{\varXi}\triangleq\bm{H}\bm{X}-\bm{S}$ is the MUI among the $M$ users. Note that the sum-rate and signal-to-interference-plus-noise ratio are typically adopted as the performance metrics in the beamforming design, while the MUI which considers the performance of all communication symbols as a whole is widely adopted as the performance metric in waveform design~\cite{bazzi2023integrated,wang2024robust}.

For radar sensing, we consider a scenario of the complex electromagnetic environment, where various jamming signals exist. In particular, the ISRJ is a widely considered jamming which can generate false targets and degrade radar detection performance. By leveraging the rectangular pulse train to sample the transmit signal~\cite{wang2007mathematic}, the interrupted sampling function is defined as
\begin{equation}
	g(t)={\rm rect}\big(\frac{t}{T_{\rm p}}\big)\sum_{n = 0}^{N_{\rm s}-1} \delta (t-n T_{\rm s}), 
\end{equation}
where $T_{\rm p}$ denotes the pulse width, $T_{\rm s}$ denotes the sampling repetition period, and $N_{\rm s}$ is the number of jamming slices. We use $\bm{g}$ to represent the interrupted sampling function in discrete time. Following~\cite{zhou2020joint}, the ISRJ signal is denoted as $\bm{\varUpsilon}=\bm{X}\odot\bm{g}$.

Note that the combiner is applied first, followed by the receive filtering since the combiner can enhance the echo signal from the detection angle while suppressing signals from other directions~\cite{fishler2006spatial}. If the combiner is applied first, the number of signals to be processed is significantly reduced, and the signal-to-noise ratio (SNR) is greatly improved, which facilitates the subsequent receive filtering. Therefore, the ISAC BS first uses $L$ combiners, with the $l$th combiner denoted as $\bm{a}(N_{\rm r},\theta_l),~\forall l\in \mathcal{L}\triangleq\{1,2,\ldots,L\}$, to filter the received signal for target detection, where
\begin{equation}
	\bm{a}(N,\theta)=\frac{1}{\sqrt{N}}\big[1,e^{j\pi\sin(\theta)},\ldots,e^{j(N-1)\pi\sin(\theta)}\big]^{\rm T},
\end{equation}
is the steering vector of $N$ antennas towards the angle $\theta$, and $\theta_l$ is the detection angle of the ISAC BS. Therefore, the received signal by the ISAC BS, denoted as $\bm{Y}_{\rm s}~\in\mathbb{C}^{N_{\rm r}\times P}$, can be expressed as
\begin{align}
	\bm{Y}_{\rm s}=&\sum_{w = 1}^{W}\alpha_w\bm{a}(N_{\rm r},\phi_w)\bm{a}^{\rm H}(N_{\rm t},\phi_w)\bm{X} \nonumber\\
	&+\alpha_0\bm{a}(N_{\rm r},\varphi)\bm{a}^{\rm H}(N_{\rm t},\varphi)\bm{\varUpsilon}+\bm{N}_{\rm s},
\end{align}
where $\alpha$ is the complex amplitude, $\phi_w$ is the angle of the $w$th target, $\varphi$ is the angle of the jammer, and $\bm{N}_{\rm s} \triangleq [\bm{n}_{\rm s,1},\ldots,\bm{n}_{\rm s,P}]\in\mathbb{C}^{N_{\rm r} \times P}$ is the AWGN, with $\bm{n}_{\rm s,p}\sim\mathcal{CN}(\bm{0},\sigma_{\rm s}^2\bm{I}_{N_{\rm r}})$. In this work, the angles of targets and the jammer are unknown. We try $L$ different detection angles $\{ \theta_l,~\forall l\in\mathcal{L} \}$ to detect the potential targets, where $L$ can be flexibly set according to the requirement of angle precision. The output signal of the $l$th combiner can be denoted as $\bm{y}_l=\big(\bm{a}^{\rm H}(N_{\rm r},\theta_l)\bm{Y}_{\rm s}\big)^{\rm T}$. After ignoring the constant terms, the potential target component and the potential jamming component can be expressed as $\bm{y}_{{\rm t},l}=\big(\bm{a}^{\rm H}(N_{\rm t},\theta_l)\bm{X}\big)^{\rm T}$ and $\bm{y}_{{\rm j},l}=\big(\bm{a}^{\rm H}(N_{\rm t},\theta_l)\bm{\varUpsilon}\big)^{\rm T},~\forall l\in \mathcal{L},$ respectively.

To detect the potential targets from $\boldsymbol{y}_l$, the receive filtering is generally employed. Denote the impulse response of the $l$th receive filter as $\bm{v}_l~\in\mathbb{C}^{P}$. Then, the receive filter output can be expressed as
\begin{equation}\label{z_l}
	\bm{z}_l=\bm{v}_l \circledast \bm{y}_l,\forall l \in \mathcal{L}.
\end{equation}
We denote the potential target component and the potential jamming component of $\bm{z}_l$ as $\bm{b}_l=\bm{v}_l \circledast \bm{y}_{{\rm t},l}$ and $\bm{d}_l=\bm{v}_l \circledast \bm{y}_{{\rm j},l},\forall l \in \mathcal{L},$ respectively. The length of $\bm{z}_l$ is $2P-1$, and the $P$th entry is the matched point. Thus, the power of the $P$th entry of $\bm{z}_l$, denoted as $\big|\bm{z}_l[P]\big|^2$, is the mainlobe level, while the power of the other $2P-2$ entries is the sidelobe levels. 

\section{Joint Transmit Waveform and Receive Filter Design}
The transmit waveform from the ISAC BS is closely related to the communication performance in terms of the MUI and radar performance in terms of target detection. In fact, the waveform design in this paper includes the beamforming design of the ISAC BS, where each column of $\boldsymbol{X}$ is a beamforming vector for a time slot. The matched filters, as the widely used receive signal processing unit for radar systems, indicate the correlation between the echo signal and the transmit signal and can be carefully designed to improve the target detection performance. In the following, we will propose the JTMD scheme using the matched filter. Moreover, since the mismatched filters offer more flexibility for the system design and can better suppress the sidelobes, we will propose the JTMMD scheme based on the JTMD scheme.

We jointly design the transmit waveform and receive filters, aiming to minimize the MUI subject to the constraints of the target mainlobe, jamming mainlobe and peak sidelobe level of the receive filter output, the transmit power of the ISAC BS, and the unit power of the receive filters. Then, the joint design problem can be formulated as
\begin{subequations}\label{problem}
\begin{align}
		\underset{\bm{X},\bm{v}}{\min}~~~& \|\bm{\varXi}\| _{\rm F}  \\
		\mathrm{s.t.} 
		~~~&\big|\bm{b}_l[P]\big|^2\geq  \zeta_l, ~\forall l\in \mathcal{L}, \label{TargetMainlobe} \\
		~~~&\big|\bm{d}_l[P]\big|^2\leq\varepsilon_l,~\forall l\in \mathcal{L},  \label{JammingMainlobe}\\
		& \big| \bm{z}_l[i] \big|^2\leq \epsilon_l,~\forall i\in\mathcal{I}\triangleq\{1,2,\ldots,2P-1,~i\neq P\}, \label{Sidelobe}\\
		&\big\| \bm{X}[:,p] \big\|_2^2\leq P_{\rm t},~\forall p\in \mathcal{P}, \label{BS_Power} \\
		&\| \bm{v}_l\|_2^2= 1,~\forall l\in \mathcal{L}, \label{FilterPower}
\end{align}
\end{subequations}
where $\zeta_l,\varepsilon_l$ and $\epsilon_l$ represent the thresholds of the target mainlobe, the jamming mainlobe, and the sidelobe, respectively. $P_{\rm t}$ denotes the maximum transmit power of the ISAC BS. Note that \eqref{TargetMainlobe} indicates the constraints of the target mainlobe and \eqref{JammingMainlobe} indicates the constraints of the jamming mainlobe. Based on \eqref{z_l}, we use \eqref{Sidelobe} to indicate the constraints of the peak sidelobe level of the receive filter output. \eqref{BS_Power} is the transmit power constraint of the ISAC BS and \eqref{FilterPower} indicates the unit power constraints of the receive filters.

\subsection{JTMD Scheme}\label{MF_section}
The matched filter, denoted as the conjugate of $\bm{y}_{{\rm t},l}$, can be expressed as
\begin{equation}\label{obtainMF}
    \bm{v}_{{\rm MF},l}\triangleq \frac{\bm{X}^{\rm H}\bm{a}(N_{\rm t},\theta_l)}{\big\|\bm{X}^{\rm H}\bm{a}(N_{\rm t},\theta_l)\big\|_2},~\forall l \in \mathcal{L}, 
\end{equation}
where the normalization is to ensure the unit power of the matched filter. We substitute $\boldsymbol{v}_l$ in~\eqref{problem} by $\boldsymbol{v}_{{\rm{MF}},l}$ and introduce \eqref{obtainMF} as an additional constraint of \eqref{problem}. However, such a problem is still non-convex. Before adopting the ADMM, we introduce auxiliary vectors $\bm{c}\triangleq [c_1,\ldots,c_L]^{\rm T}\in\mathbb{C}^{L} ,\bm{q}\triangleq [q_1,\ldots,q_L]^{\rm T}\in\mathbb{C}^{L},$ and $\bm{\gamma}\triangleq \big[\bm{\gamma}_1^{\rm T},\ldots,\bm{\gamma}_L^{\rm T}\big]^{\rm T}\in\mathbb{C}^{(2P-1)L} $. Then,~\eqref{problem} can be transformed to
\begin{align}\label{problem1}
	\underset{\bm{X},\bm{v}}{\min}~~& \|\bm{\varXi}\| _{\rm F} \nonumber\\
	~~\mathrm{s.t.} ~~&  \bm{b}_l[P] = c_l,~\bm{d}_l[P]=q_l,~\forall l\in \mathcal{L}, \nonumber \\
	& | c_l|^2 \geq \zeta_l,~ | q_l|^2 \leq \varepsilon _l,~\forall l \in \mathcal{L}, \nonumber\\
	& \bm{z}_l=\bm{\gamma}_l,~ \big|\bm{\gamma}_l[i]\big|^2 \leq \epsilon_l,~\forall i \in \mathcal{I},~\forall l \in \mathcal{L}, \nonumber\\
	&\big\| \bm{X}[:,p] \big\|_2^2\leq P_{\rm t},~\forall p\in \mathcal{P}, \nonumber\\
	&\bm{v}_l=\bm{v}_{{\rm MF},l},~\forall l \in \mathcal{L},
\end{align}
where $\bm{v}\triangleq\big[\bm{v}_1^{\rm T},\ldots,\bm{v}_L^{\rm T}\big]^{\rm T}~\in\mathbb{C}^{PL}$. In the JTMD scheme, the receive filter $\boldsymbol{v}_{l}$ is specified as the matched filter $\boldsymbol{v}_{{\rm MF},l}$ through $\bm{v}_l=\bm{v}_{{\rm MF},l}$ in~\eqref{problem1} and is fixed once the transmit waveform $\boldsymbol{X}$ is given. However, in this way,~\eqref{problem1} becomes a non-convex problem with respect to $\bm{X}$ and therefore is difficult to solve. To deal with this problem, we retain $\boldsymbol{v}_{l}$ and leverage the augmented Lagrangian function to relax the equality constraint $\bm{v}_l=\bm{v}_{{\rm MF},l}$. The augmented Lagrangian function of this problem is
\begin{flalign}\label{Lagrangian1}
	&L_{\rho_1}(\bm{X},\bm{v},\bm{c},\bm{q},\bm{\gamma},\bm{\kappa},\bm{\tau},\bm{\omega},\bm{\lambda})\nonumber\\
	&=\|\bm{\varXi}\| _{\rm F}+\sum_{l = 1}^{L} \Bigg(\frac{\rho_1}{2}\Big|\bm{b}_l[P]- c_l+\frac{\kappa_l}{\rho_1}\Big|^2 \nonumber\\
	&~~~+\frac{\rho_1}{2}\Big|\bm{d}_l[P]- q_l+\frac{\tau_l}{\rho_1}\Big|^2+\frac{\rho_1}{2}\Big\|\bm{z}_l-\bm{\gamma}_l+\frac{\bm{\omega}_l}{\rho_1}\Big\|_2^2 \nonumber\\
	&~~~+\frac{\rho_1}{2}\Big\|\bm{v}_l-\bm{v}_{{\rm MF},l}+\frac{\bm{\lambda}_l}{\rho_1}\Big\|_2^2\Bigg),& 
\end{flalign}
where $\bm{\kappa}\triangleq [\kappa_1,\ldots,\kappa_L]^{\rm T}\in\mathbb{C}^{L},\bm{\tau}\triangleq [\tau_1,\ldots,\tau_L]^{\rm T}\in\mathbb{C}^{L},\bm{\omega}\triangleq \big[\bm{\omega}_1^{\rm T},\ldots,\bm{\omega}_L^{\rm T}\big]^{\rm T}\in\mathbb{C}^{(2P-1)L},$ and $\bm{\lambda}\triangleq\big[\bm{\lambda}_1^{\rm T},\ldots,\bm{\lambda}_L^{\rm T}\big]^{\rm T}\in\mathbb{C}^{PL}$ are the dual variables, and $\rho_1>0$ is a penalty parameter. The variables are updated alternately following the order $\{ \bm{X},\bm{v},\bm{c},\bm{q},\bm{\gamma},\bm{\kappa},\bm{\tau},\bm{\omega},\bm{\lambda}\} $ according to the convention of ADMM.

We denote $k$ as the iteration counter. Then, at the $(k+1)$th iteration, we update $\bm{X}^{(k+1)}$ by solving
\begin{align}\label{X}
	\underset{\bm{X}}{\min}~&\|\bm{\varXi}\| _{\rm F}+\sum_{l = 1}^{L} \Bigg(\frac{\rho_1}{2}\Big|\bm{b}_l[P]- c_l+\frac{\kappa_l}{\rho_1}\Big|^2 \nonumber\\
	&+\frac{\rho_1}{2}\Big|\bm{d}_l[P]- q_l+\frac{\tau_l}{\rho_1}\Big|^2+\frac{\rho_1}{2}\Big\|\bm{z}_l-\bm{\gamma}_l+\frac{\bm{\omega}_l}{\rho_1}\Big\|_2^2\Bigg) \nonumber\\
	~~\mathrm{s.t.} ~&\big\| \bm{X}[:,p] \big\|_2^2\leq P_{\rm t},~\forall p \in \mathcal{P}.
\end{align}

The updating of $\bm{v}$ can be decomposed into $L$ independent subproblems and solved in parallel as
\begin{align}\label{v}
	\bm{v}^{(k+1)}_l=&{\rm arg}\underset{\bm{v}_l}{\min}~\frac{\rho_1}{2}\Big|\bm{b}_l[P]-c_l+\frac{\kappa_l}{\rho_1}\Big|^2 \nonumber\\
	&+\frac{\rho_1}{2}\Big|\bm{d}_l[P]- q_l+\frac{\tau_l}{\rho_1}\Big|^2+\frac{\rho_1}{2}\Big\|\bm{z}_l-\bm{\gamma}_l+\frac{\bm{\omega}_l}{\rho_1}\Big\|_2^2 \nonumber\\
	&+\frac{\rho_1}{2}\Big\|\bm{v}_l-\bm{v}_{{\rm MF},l}+\frac{\bm{\lambda}_l}{\rho_1}\Big\|_2^2,~\forall l \in \mathcal{L}.
\end{align}

Note that~\eqref{X} and~\eqref{v} are convex and can be computed using the CVX toolbox. Through~\eqref{v}, we can find that $\bm{v}_l$ does not always equal $\bm{v}_{{\rm MF},l}$ during the optimization. However, the primal feasibility and the dual feasibility of ADMM eventually ensure that after the algorithm converges, $\bm{v}_l=\bm{v}_{{\rm MF},l},~\forall l \in \mathcal{L}$. Then, after omitting the terms independent of $\bm{c}$, we can update $\bm{c}$ by solving
\begin{align}
\underset{\bm{c}}{\min}~~& \sum_{l = 1}^{L} \frac{\rho_1}{2}\Big|\bm{b}_l[P]-c_l+\frac{\kappa_l}{\rho_1}\Big|^2 \nonumber\\
~~\mathrm{s.t.}~~&|c_l|^2\geq \zeta_l,~\forall l \in \mathcal{L}.
\end{align}
By setting the gradient of $c_l$ to zero, we can obtain
\begin{equation}
	\begin{aligned}\label{c}
		c_l&=
	\begin{cases}
		\bm{b}_l[P]+\kappa_l/\rho_1,&{\rm if}~|c_l|^2\geq \zeta_l,\\
        \sqrt{\zeta_l} \frac{\bm{b}_l[P]+\kappa_l/\rho_1}{\big|\bm{b}_l[P]+\kappa_l/\rho_1\big|},&{\rm otherwise},
    \end{cases}
	~\forall l \in \mathcal{L}.
\end{aligned}
\end{equation}

Similarly, $\bm{q}$ and $\bm{\gamma}$ can be obtained respectively by
\begin{equation}
	\begin{aligned}\label{QandGamma}
		q_l&=
	\begin{cases}
		\bm{d}_l[P]+\tau_l/\rho_1,&{\rm if}~|q_l|^2\leq \varepsilon_l,\\
        \sqrt{\varepsilon_l} \frac{\bm{d}_l[P]+\tau_l/\rho_1}{\big|\bm{d}_l[P]+\tau_l/\rho_1\big|},&{\rm otherwise},
    \end{cases}
	~\forall l \in \mathcal{L},\\
		\bm{\gamma}_l[i]&=
	\begin{cases}
		\bm{\eta}_l[i],&{\rm if}~\big|\bm{\gamma}_l[i]\big|^2 \leq \epsilon_l,\\
        \sqrt{\epsilon_l} \frac{\bm{\eta}_l[i]}{\big|\bm{\eta}_l[i]\big|},&{\rm otherwise},
    \end{cases}
	~\forall i \in \mathcal{I},~\forall l \in \mathcal{L},
\end{aligned}
\end{equation}
where $\bm{\eta}_l\triangleq \bm{z}_l+\bm{\omega}_l/\rho_1,~\forall l \in \mathcal{L}$. Then, the dual variables are updated as
\begin{align}\label{dual_variables}
\kappa_l^{(k+1)}&=\kappa_l^{(k)}+\rho_1\big(\bm{b}_l[P]-c_l\big),~\forall l \in \mathcal{L},\nonumber\\
\tau_l^{(k+1)}&=\tau_l^{(k)}+\rho_1\big(\bm{d}_l[P]-q_l\big),~\forall l \in \mathcal{L},\nonumber\\
\bm{\omega}_l^{(k+1)}&=\bm{\omega}_l^{(k)}+\rho_1\big(\bm{z}_l-\bm{\gamma}_l\big),~\forall l \in \mathcal{L},\nonumber\\
\bm{\lambda}_l^{(k+1)}&=\bm{\lambda}_l^{(k)}+\rho_1(\bm{v}_l-\bm{v}_{{\rm MF},l}),~\forall l \in \mathcal{L}.
\end{align}
The complete procedures of the JTMD scheme are summarized in \textbf{Algorithm~\ref{alg1}}. The maximum number of iterations is denoted as $K$. We initialize the transmit waveform as the orthogonal LFM waveform to prioritize sensing requirements and ensure robust sensing performance even under jamming conditions~\cite{zhong2024p}, which can be denoted as
\begin{equation}\label{O_LFM}
	\bm{X}_{\rm s}[m,p]=\frac{1}{\sqrt{N_{\rm t} P}}e^{j2\pi m(p-1)/P} e^{j\pi (p-1)^2/P}.
\end{equation}

The bandwidth of~\eqref{O_LFM} is normalized by sampling the standard LFM waveform and setting the sampling interval as the inverse of the signal bandwidth. In fact, the number of targets, as well as the range and angles of each target, can also be estimated. If there exists a target in the detection angle $\theta_l$, a peak will occur in $|\bm{z}_l |$. Therefore, we can estimate $W$ as the total number of peaks of $| \bm{z}_l |$. Then we can estimate $\phi_w$ as $\theta_l$. Based on the locations of the peaks of $| \bm{z}_l |,~\forall l \in \mathcal{L}$, we can estimate the round-trip delay which essentially corresponds to the range of each target according to the principle of radar ranging in~\cite[Ch 7.2]{richards2005fundamentals}.


\begin{algorithm}[!t]
	\caption{JTMD Scheme}
	\label{alg1}
	\begin{algorithmic}[1]
		\REQUIRE $\bm{H}, \bm{S}, \bm{g}$.
		\ENSURE $\bm{X}$, $\bm{v}$.
		\STATE Set $\bm{X}^{(0)}$, $\bm{v}^{(0)}$ with $\bm{X}_{\rm s}$ by~\eqref{obtainMF}.
		\STATE Set $\bm{c}^{(0)},\bm{q}^{(0)},\bm{\gamma}^{(0)},\bm{\kappa}^{(0)},\bm{\tau}^{(0)},\bm{\omega}^{(0)}$ and $\bm{\lambda}^{(0)}$ zero.
		\STATE Set $k\leftarrow 0$.
		\WHILE{$k < K$}
		\STATE $k\leftarrow k+1$.
		\STATE Update $\bm{X}^{(k)}$ by solving~\eqref{X}.
		\STATE Update $\bm{v}^{(k)}$ by solving~\eqref{v}.
		\STATE Update $\bm{c}^{(k)},\bm{q}^{(k)}$ and $\bm{\gamma}^{(k)}$ by~\eqref{c} and \eqref{QandGamma}.
		\STATE Update $\bm{\kappa}^{(k)},\bm{\tau}^{(k)},\bm{\omega}^{(k)}$ and $\bm{\lambda}^{(k)}$ by~\eqref{dual_variables}.
		\ENDWHILE
		\STATE $\bm{X}\leftarrow \bm{X}^{(K)},\bm{v}\leftarrow \bm{v}^{(K)}$.
	\end{algorithmic}
\end{algorithm}

\subsection{JTMMD Scheme}
Based on the JTMD scheme, we further consider the JTMMD scheme, since the mismatched filters can better suppress the jamming and sidelobe levels. By introducing auxiliary vectors $\bm{e}\triangleq [e_1,\ldots,e_L]^{\rm T}\in\mathbb{C}^{L},\bm{r}\triangleq [r_1,\ldots,r_L]^{\rm T}\in\mathbb{C}^{L},\bm{\xi}\triangleq\big[\bm{\xi}_1^{\rm T},\ldots,\bm{\xi}_L^{\rm T}\big]^{\rm T}\in\mathbb{C}^{(2P-1)L}$, and $\bm{\beta}\triangleq\big[\bm{\beta}_1^{\rm T},\ldots,\bm{\beta}_L^{\rm T}\big]^{\rm T}\in\mathbb{C}^{PL}$, we can transform~\eqref{problem} into
\begin{align}\label{problem2}
	\underset{\bm{X},\bm{v}}{\min}~~& \|\bm{\varXi}\| _{\rm F} \nonumber\\
	~~\mathrm{s.t.} ~~&  \bm{b}_l[P] = e_l,\bm{d}_l[P]=r_l,~\forall l\in \mathcal{L}, \nonumber \\
	& | e_l|^2 \geq \zeta_l,~ | r_l|^2 \leq \varepsilon _l,~\forall l \in \mathcal{L}, \nonumber\\
	& \bm{z}_l=\bm{\xi}_l,~\big|\bm{\xi}_l[i]\big|^2 \leq \epsilon_l,~\forall i \in \mathcal{I},~\forall l \in \mathcal{L}, \nonumber\\
	&\big\| \bm{X}[:,p] \big\|_2^2\leq P_{\rm t},~\forall p\in \mathcal{P}, \nonumber\\
	&\bm{v}_l=\bm{\beta}_l,~\big\| \bm{\beta}_l\big\|_2^2= 1,~\forall l\in \mathcal{L}.
\end{align}

We still adopt the ADMM to solve this problem. The augmented Lagrangian function of~\eqref{problem2} is
\begin{flalign}\label{Lagrangian2}
	&L_{\rho_2}(\bm{X}, \bm{v},\bm{e},\bm{r},\bm{\xi},\bm{\beta},\bm{\varpi},\bm{\varrho},\bm{\mu},\bm{\varsigma})\nonumber\\
	&=\|\bm{\varXi}\| _{\rm F}+\sum_{l = 1}^{L} \Bigg(\frac{\rho_2}{2}\Big|\bm{b}_l[P]- e_l+\frac{\varpi_l}{\rho_2}\Big|^2 \nonumber\\
	&~~~+\frac{\rho_2}{2}\Big|\bm{d}_l[P]- r_l+\frac{\varrho_l}{\rho_2}\Big|^2+\frac{\rho_2}{2}\Big\|\bm{z}_l-\bm{\xi}_l+\frac{\bm{\mu}_l}{\rho_2}\Big\|_2^2 \nonumber\\
	&~~~+\frac{\rho_2}{2}\Big\|\bm{v}_l-\bm{\beta}_l+\frac{\bm{\varsigma}_l}{\rho_2}\Big\|_2^2\Bigg),&
\end{flalign}
where $\bm{\varpi}\triangleq [\varpi_1,\ldots,\varpi_L]^{\rm T}\in\mathbb{C}^{L},\bm{\varrho}\triangleq [\varrho_1,\ldots,\varrho_L]^{\rm T}\in\mathbb{C}^{L},\bm{\mu}\triangleq\big[\bm{\mu}_1^{\rm T},\ldots,\bm{\mu}_L^{\rm T}\big]^{\rm T}\in\mathbb{C}^{(2P-1)L},$ and $\bm{\varsigma}\triangleq\big[\bm{\varsigma}_1^{\rm T},\ldots,\bm{\varsigma}_L^{\rm T}\big]^{\rm T}\in\mathbb{C}^{PL}$ are the dual variables, and $\rho_2>0$ is a penalty parameter. The variables are updated alternately following the order $\{ \bm{X}, \bm{v},\bm{e},\bm{r},\bm{\xi},\bm{\beta},\bm{\varpi},\bm{\varrho},\bm{\mu},\bm{\varsigma}\} $.

Since~\eqref{Lagrangian2} and~\eqref{Lagrangian1} have the same structure, we can directly use \textbf{Algorithm~\ref{alg1}} to obtain $\bm{X},\bm{v},\bm{e},\bm{r}$ and $\bm{\xi}$ by replacing $[\bm{c},\bm{q},\bm{\gamma},\bm{\lambda},\bm{v}_{{\rm MF},l}]$ with $[\bm{e},\bm{r},\bm{\xi},\bm{\mu},\bm{\beta}_l]$. After omitting the terms irrelevant to $\bm{\beta}$, we update $\bm{\beta}$ by
\begin{algorithm}[!t]
	\caption{JTMMD Scheme}
	\label{alg2}
	\begin{algorithmic}[1]
		\REQUIRE $\bm{H}, \bm{S}, \bm{g}$.
		\ENSURE $\bm{X}$, $\bm{v}$.
		\STATE Set $\bm{X}^{(0)}$, $\bm{v}^{(0)}$ with $\bm{X}_{\rm s}$ by~\eqref{obtainMF}.
		\STATE Set $\bm{e}^{(0)},\bm{r}^{(0)},\bm{\xi}^{(0)},\bm{\beta}^{(0)},\bm{\varpi}^{(0)},\bm{\varrho}^{(0)},\bm{\mu}^{(0)}$ and $\bm{\varsigma}^{(0)}$ zero.
		\STATE Set $k\leftarrow 0$.
		\WHILE{$k<K$}
		\STATE $k\leftarrow k+1$.
		\STATE Update $\bm{X}^{(k)},\bm{v}^{(k)},\bm{e}^{(k)},\bm{r}^{(k)},\bm{\xi}^{(k)}$ by \textbf{Algorithm~\ref{alg1}}.
		\STATE Update $\bm{\beta}^{(k)}$ by~\eqref{Beta}.
		\STATE Update $\bm{\varpi}^{(k)},\bm{\varrho}^{(k)},\bm{\mu}^{(k)}$ and $\bm{\varsigma}^{(k)}$ by~\eqref{dual_variables2}.
		\ENDWHILE
		\STATE $\bm{X}\leftarrow \bm{X}^{(K)},\bm{v}\leftarrow \bm{v}^{(K)}$.
	\end{algorithmic}
\end{algorithm}
\begin{align}
\underset{\bm{\beta}}{\min}~~~& \sum_{l = 1}^{L} \frac{\rho_2}{2}\Big\|\bm{v}_l-\bm{\beta}_l+\frac{\bm{\varsigma}_l}{\rho_2}\Big\|_2^2 \nonumber\\
~~~\mathrm{s.t.} ~~~&\big\|\bm{\beta}_l\big\|_2^2=1,~\forall l\in \mathcal{L}.
\end{align}

By setting the gradient of $\bm{\beta}_l$ to zero, we can obtain
\begin{equation}\label{Beta}
	\bm{\beta}_l=\frac{\bm{v}_l+\bm{\mu}_l/\rho_2}{\|\bm{v}_l+\bm{\mu}_l/\rho_2\|_2},~\forall l \in \mathcal{L}. 
\end{equation}

The dual variables are updated as
\begin{align}\label{dual_variables2}
\varpi_l^{(k+1)}&=\varpi_l^{(k)}+\rho_2\big(\bm{b}_l[P]-e_l\big),~\forall l \in \mathcal{L},\nonumber\\
\varrho_l^{(k+1)}&=\varrho_l^{(k)}+\rho_2\big(\bm{d}_l[P]-r_l\big),~\forall l \in \mathcal{L},\nonumber\\
\bm{\mu}_l^{(k+1)}&=\bm{\mu}_l^{(k)}+\rho_2\big(\bm{z}_l-\bm{\xi}_l\big),~\forall l \in \mathcal{L}, \nonumber\\
\bm{\varsigma}_l^{(k+1)}&=\bm{\varsigma}_l^{(k)}+\rho_2\big(\bm{v}_l-\bm{\beta}_l\big),~\forall l \in \mathcal{L}.
\end{align}

The complete procedures of the JTMMD scheme are summarized in \textbf{Algorithm~\ref{alg2}}.

Note that both the JTMD and JTMMD schemes can be readily extended to wideband scenarios, e.g., extending to the OFDM framework with the wideband channel modeling~\cite{zhang2024transmit}.
\section{Simulation Results}\label{sec.simulation}
To evaluate the system performance, we consider an ISAC system, where an ISAC BS serves $M=2$ communication users and senses $W=2$ targets simultaneously. The ISAC BS is equipped with $N_{\rm t}=32$ transmit antennas and $N_{\rm r}=16$ receive antennas. The ISAC BS uses $L=8$ combiners to detect the angle space of interest and the maximum transmit power of the ISAC BS is $P_{\rm t}=2{\rm W}$. The  pulse width and the sampling repetition period of the jammer are set to be $T_{\rm p}=2.5{\rm us}$ and $T_{\rm s}=7.5{\rm us}$, respectively. 

\begin{figure}[!t]
	\centerline{\includegraphics[height=6 cm]{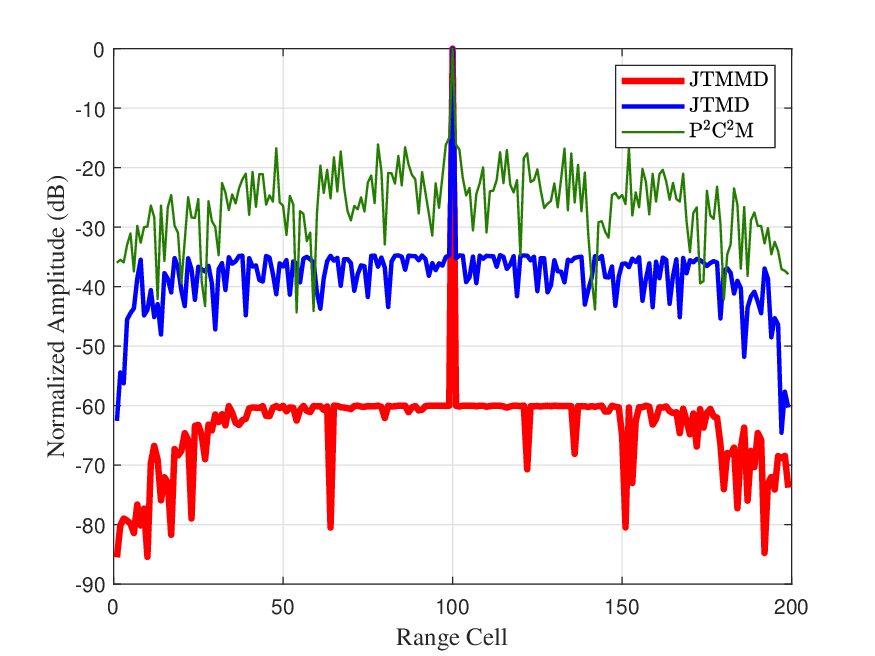}}
	\caption{Comparisons of the sidelobe levels under the same mainlobe level.}
	\label{PC}
\end{figure}

For fair comparisons, we extend the existing scheme named parallel product complex circle manifold method (${\rm P}^2{\rm C}^2{\rm M}$)~\cite{zhong2024p} to the same scenario. By replacing the non-convex constant modulus constraint with the convex transmit power constraint in~\eqref{problem}, we can modify the original problem considered by ${\rm P}^2{\rm C}^2{\rm M}$ and adapt ${\rm P}^2{\rm C}^2{\rm M}$ for our problem. In addition, since the transmit power constraint can provide more degrees of freedom for the system design than the constant modulus constraint, ${\rm P}^2{\rm C}^2{\rm M}$ can achieve better performance in the considered scenario than in~\cite{zhong2024p}. We first compare the communication performance in terms of the MUI under SNR of 10 dB. The MUI of JTMD and JTMMD is less than $10^{-5}$ and the MUI of ${\rm P}^2{\rm C}^2{\rm M}$ is less than $10^{-2}$. In other words, both the proposed schemes and ${\rm P}^2{\rm C}^2{\rm M}$ can achieve satisfactory MUI. Therefore, we focus on the comparisons of radar performance. As shown in Fig.~\ref{PC}, we compare the mainlobe level and sidelobe levels of the receive filter output. The ratio of the mainlobe level over the peak sidelobe level of ${\rm P}^2{\rm C}^2{\rm M}$, JTMD, and JTMMD is around 15 dB, 35 dB and 60 dB, respectively. Both the proposed schemes outperform ${\rm P}^2{\rm C}^2{\rm M}$ with lower sidelobe levels under the same mainlobe level, demonstrating the effectiveness of the proposed schemes.

\begin{figure}[!t]
	\centerline{\includegraphics[height=6 cm]{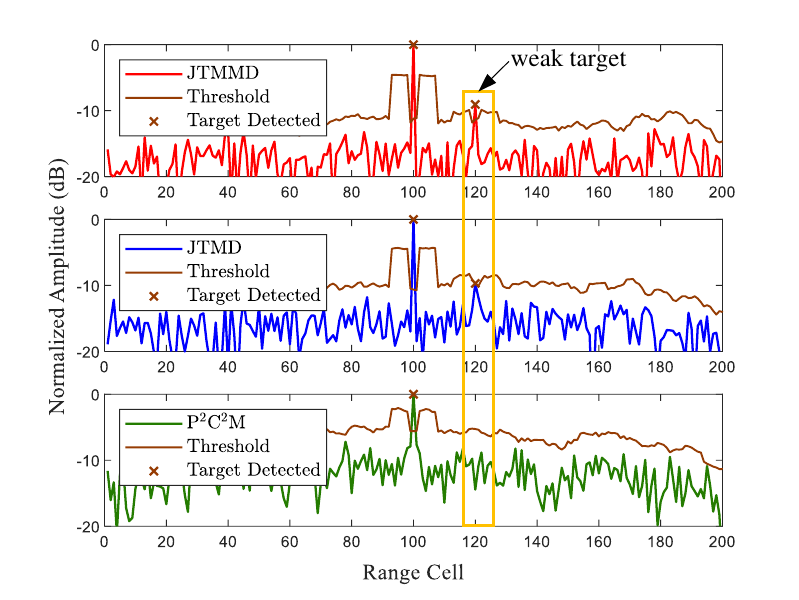}}
	\caption{Comparisons of weak target detection for different schemes.}
	\label{WeakTarget}
\end{figure}

\begin{figure}[!t]
	\centerline{\includegraphics[height=6 cm]{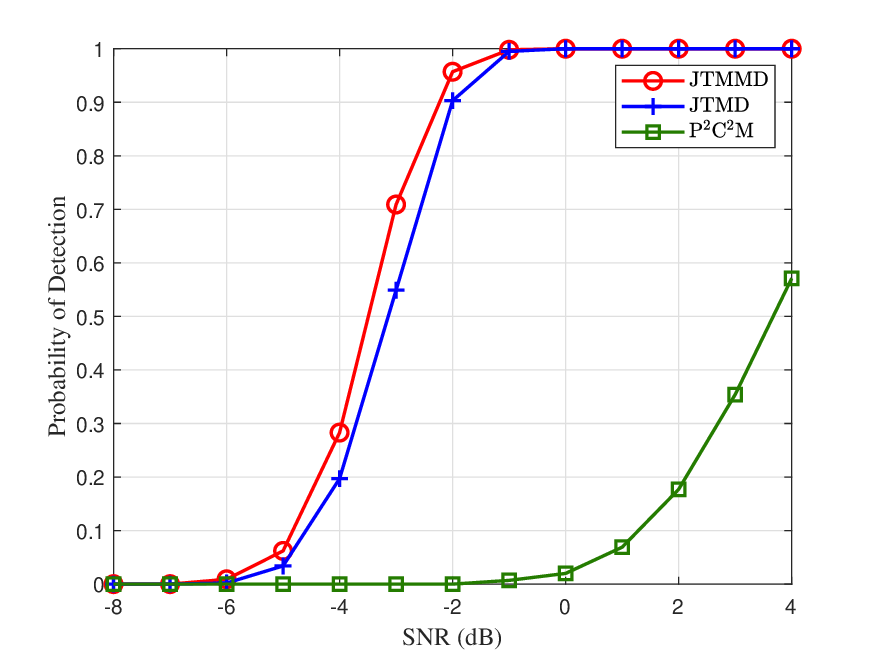}}
	\caption{Comparisons of the probability of detection versus SNRs.}
	\label{Pd}
\end{figure}

To illustrate the target detection performance, we consider the situation of detecting two targets, one of which is a strong target located at the $100$th range cell while the other one is a weak target located at the $120$th range cell~\cite{du2024reshaping}. We utilize the constant false alarm rate detector for target detection with the probability of false alarm fixed to be $10^{-5}$. The sensing SNR, defined as the power ratio of the echo signal over the noise, is 10 dB. As shown in Fig.~\ref{WeakTarget}, both the proposed schemes can successfully detect two targets. ${\rm P}^2{\rm C}^2{\rm M}$ fails to detect the weak target, since the weak target is overwhelmed by the jamming and noise. In particular, the threshold to detect the weak target for JTMMD is lower than that for JTMD, indicating that the former achieves better performance of target detection than the latter.

To evaluate the average detection performance, we show the probability of detection under different SNR levels through 1000 Monte Carlo trials in Fig.~\ref{Pd}. Both the proposed schemes achieve much better target detection performance than ${\rm P}^2{\rm C}^2{\rm M}$. Moreover, JTMMD outperforms JTMD, illustrating that the mismatched filter achieves better performance for target detection with jamming than the matched filter.

\section{Conclusion}
In this paper, we have proposed the JTMD and JTMMD schemes for the ISAC system to suppress jamming, where the number of the targets as well as the range and angles of each target can also be estimated. Simulation results have shown that both the JTMD and JTMMD schemes can achieve superior performance in terms of communication MUI and radar detection performance. In the future, we will explore other advanced jamming suppression methods as well as symbol-level design methods for the ISAC system.
\bibliographystyle{IEEEtran}
\bibliography{IEEEabrv,mybib}

\begin{thebibliography}{10}
\providecommand{\url}[1]{#1}
\csname url@samestyle\endcsname
\providecommand{\newblock}{\relax}
\providecommand{\bibinfo}[2]{#2}
\providecommand{\BIBentrySTDinterwordspacing}{\spaceskip=0pt\relax}
\providecommand{\BIBentryALTinterwordstretchfactor}{4}
\providecommand{\BIBentryALTinterwordspacing}{\spaceskip=\fontdimen2\font plus
\BIBentryALTinterwordstretchfactor\fontdimen3\font minus
  \fontdimen4\font\relax}
\providecommand{\BIBforeignlanguage}[2]{{%
\expandafter\ifx\csname l@#1\endcsname\relax
\typeout{** WARNING: IEEEtran.bst: No hyphenation pattern has been}%
\typeout{** loaded for the language `#1'. Using the pattern for}%
\typeout{** the default language instead.}%
\else
\language=\csname l@#1\endcsname
\fi
#2}}
\providecommand{\BIBdecl}{\relax}
\BIBdecl

\bibitem{wei2023integrated}
Z.~Wei \emph{et~al.}, ``Integrated sensing and communication signals toward
  5{G}-{A} and 6{G}: A survey,'' \emph{IEEE Internet Things J.}, vol.~10,
  no.~13, pp. 11\,068--11\,092, Jan. 2023.

\bibitem{liu2022integrated}
F.~Liu \emph{et~al.}, ``Integrated sensing and communications: Toward
  dual-functional wireless networks for 6{G} and beyond,'' \emph{IEEE J. Sel.
  Areas Commun.}, vol.~40, no.~6, pp. 1728--1767, June 2022.

\bibitem{kaushik2024toward}
A.~Kaushik \emph{et~al.}, ``Toward integrated sensing and communications for
  6{G}: Key enabling technologies, standardization, and challenges,''
  \emph{IEEE Commun. Standards Mag.}, vol.~8, no.~2, pp. 52--59, June 2024.

\bibitem{meng2023throughput}
K.~Meng, Q.~Wu, S.~Ma, W.~Chen, K.~Wang, and J.~Li, ``Throughput maximization
  for {UAV}-enabled integrated periodic sensing and communication,'' \emph{IEEE
  Trans. Wireless Commun.}, vol.~22, no.~1, pp. 671--687, Jan. 2023.

\bibitem{bazzi2023integrated}
A.~Bazzi and M.~Chafii, ``On integrated sensing and communication waveforms
  with tunable {PAPR},'' \emph{IEEE Trans. Wireless Commun.}, vol.~22, no.~11,
  pp. 7345--7360, Nov. 2023.

\bibitem{liu2021dual}
R.~Liu, M.~Li, Q.~Liu, and A.~L. Swindlehurst, ``Dual-functional
  radar-communication waveform design: A symbol-level precoding approach,''
  \emph{IEEE J. Sel. Topics Signal Process.}, vol.~15, no.~6, pp. 1316--1331,
  Sep. 2021.

\bibitem{wang2024robust}
S.~Wang, W.~Dai, H.~Wang, and G.~Y. Li, ``Robust waveform design for integrated
  sensing and communication,'' \emph{IEEE Trans. Signal Process.}, vol.~72, pp.
  3122--3138, June 2024.

\bibitem{wang2007mathematic}
X.~Wang, J.~Liu, W.~Zhang, Q.~Fu, Z.~Liu, and X.~Xie, ``Mathematic principles
  of interrupted-sampling repeater jamming {(ISRJ)},'' \emph{Sci. China Inf.
  Sci.}, vol.~50, pp. 113--123, Feb. 2007.

\bibitem{cao2021optimal}
F.~Cao, Z.~Chen, X.~Feng, C.~He, and J.~Xu, ``Optimal design of
  anti-interrupted sampling repeater jamming waveform for missile-borne radar
  based on an improved genetic algorithm,'' \emph{IET Signal Process.},
  vol.~15, no.~9, pp. 622--632, July 2021.

\bibitem{chen2019band}
J.~Chen, W.~Wu, S.~Xu, Z.~Chen, and J.~Zou, ``Band pass filter design against
  interrupted-sampling repeater jamming based on time-frequency analysis,''
  \emph{IET Radar Sonar Nav.}, vol.~13, no.~10, pp. 1646--1654, Oct. 2019.

\bibitem{zhou2020joint}
K.~Zhou, D.~Li, Y.~Su, and T.~Liu, ``Joint design of transmit waveform and
  mismatch filter in the presence of interrupted sampling repeater jamming,''
  \emph{IEEE Signal Process. Lett.}, vol.~27, pp. 1610--1614, Sep. 2020.

\bibitem{gao2023joint}
Y.~Gao, H.~Fan, L.~Ren, Z.~Liu, Q.~Liu, and E.~Mao, ``Joint design of waveform
  and mismatched filter for interrupted sampling repeater jamming
  suppression,'' \emph{IEEE Trans. Aerosp. Electron. Syst.}, vol.~59, no.~6,
  pp. 8037 -- 8050, Dec. 2023.

\bibitem{wang2022complementary}
F.~Wang, N.~Li, C.~Pang, C.~Li, Y.~Li, and X.~Wang, ``Complementary sequences
  and receiving filters design for suppressing interrupted sampling repeater
  jamming,'' \emph{IEEE Geosci. Remote Sens. Lett.}, vol.~19, pp. 1--5, Mar.
  2022.

\bibitem{fishler2006spatial}
E.~Fishler, A.~Haimovich, R.~S. Blum, L.~J. Cimini, D.~Chizhik, and R.~A.
  Valenzuela, ``Spatial diversity in radars—models and detection
  performance,'' \emph{IEEE Trans. Signal Process.}, vol.~54, no.~3, pp.
  823--838, Mar. 2006.

\bibitem{zhong2024p}
K.~Zhong \emph{et~al.}, ``{P2C2M}: Parallel product complex circle manifold for
  {RIS}-aided {ISAC} waveform design,'' \emph{IEEE Trans. Cogn. Commun. Netw.},
  vol.~10, no.~4, pp. 1441--1451, Aug. 2024.

\bibitem{richards2005fundamentals}
M.~A. Richards, \emph{Fundamentals of Radar Signal Processing}.\hskip 1em plus
  0.5em minus 0.4em\relax New York, NY, USA: McGraw-Hill, 2014.

\bibitem{zhang2024transmit}
X.~Zhang, X.~Wang, H.~So, A.~M. Zoubir, J.~A. Zhang, and Y.~J. Guo, ``Transmit
  waveform design for integrated wideband {MIMO} radar and bi-directional
  communications,'' \emph{IEEE Trans. Veh. Technol.}, vol.~73, no.~9, pp.
  13\,482--13\,497, Sep. 2024.

\bibitem{du2024reshaping}
Z.~Du, F.~Liu, Y.~Xiong, T.~X. Han, Y.~C. Eldar, and S.~Jin, ``Reshaping the
  {ISAC} tradeoff under {OFDM} signaling: A probabilistic constellation shaping
  approach,'' \emph{IEEE Trans. Signal Process.}, vol.~72, pp. 4782--4797, Sep.
  2024.

\end{thebibliography}
\end{document}